\documentclass[a4paper,12pt]{article}
\usepackage{amsmath,amssymb,amsfonts,amsthm}
\newtheorem{theorem}{Theorem}[section]
\newtheorem{lemma}[theorem]{Lemma}

\newtheorem{corollary}[theorem]{Corollary}

\newcommand{\dv}{\, d\mu}

\newcommand{\Rt}{\mathbb{R}^3}

\newcommand{\mf}{\mathcal{M}}

\newcommand{\vv}{\alpha}

\newcommand{\vY}{y}

\newcommand{\vYb}{\bar y}

\newcommand{\vh}{\varphi}

\newcommand{\cbn}[1]{||#1||_{{C^{1}_\beta}(\Omega)}}

\newcommand{\bs}{\mathcal{B}}
\newcommand{\hs}{\mathcal{H}}

\newcommand{\bn}[1]{||#1||_{\bs}}
\newcommand{\hn}[1]{||#1||_{\hs}}

\newcommand{\dy}{\Omega_{\rho_0}}

\title{Proof of the (local) angular momemtum-mass inequality for
  axisymmetric black holes} 
\author{Sergio Dain\\
  Max-Planck-Institut f\"ur Gravitationsphysik\\
  Am M\"uhlenberg 1\\
  14476 Golm\\
  Germany}

\begin{document}
\maketitle

\begin{abstract}
  We prove that for any vacuum, maximal, asymptotically flat,
  axisymmetric initial data for Einstein equations close to 
  extreme Kerr data, the inequality $\sqrt{J} \leq m$ is satisfied,
  where $m$ and $J$ are the total mass and angular momentum of the
  data. The proof consists in showing that extreme Kerr is a local
  minimum of the mass.
\end{abstract}

\section{Introduction}
\label{sec:introduction}
Physical arguments led to the conjecture that for every axisymmetric,
vacuum, asymptotically flat, complete, initial data set for Einstein
equations the following inequality should hold
\begin{equation}
  \label{eq:14a}
\sqrt{|J|}\leq m,
\end{equation}
where $m$ is the mass of the data and $J$ the angular momentum in the
asymptotic region.  Moreover, the equality in \eqref{eq:14a} should
imply that the data is an slice of the extreme Kerr black hole.  For a
more detailed discussion of the motivations and relevance of
\eqref{eq:14a} and related inequalities see \cite{Friedman82},
\cite{Horowitz84}, \cite{Dain02c} \cite{Dain05c} and \cite{Dain05e}.

In \cite{Dain05c} the proof of \eqref{eq:14a}, for maximal data,  was  reduced
to a  variational problem. In this article we will prove that this
variational problem has a local minimum at the extreme Kerr solution,
and hence we prove that inequality  \eqref{eq:14a} is satisfied for
axisymmetric, maximal, vacuum   data sufficiently close to extreme
Kerr data. 

We first present the mathematical  formulation of the variational
problem. 
Let $\rho,z,\phi$ be cylindrical
coordinates in $\Rt$, and let $v,Y$ two functions which depend only on
$\rho,z$.  Consider the functional defined in \cite{Dain05c}
\footnote{We have slightly changed the notation in \cite{Dain05c}: we
  rescale $v\to 4v$.}
\begin{equation}
  \label{eq:5c}
 \mf(v,Y)= \frac{1}{32\pi}\int_{\mathbb{R}^3}
  \left(|\partial  v|^2  + \rho^{-4} e^{-2v} |\partial Y |^2  \right) \dv, 
\end{equation}
where $\dv=\rho \,dz d\rho d\phi$ is the volume element in $\Rt$ and
$\partial$ denotes partial derivative with respect to $\rho$ and $z$;
that is $|\partial v|^2= v^2_{,z} +v^2_{,\rho}$.
Let $v_0$ and $Y_0$ denote the extreme Kerr initial data given
explicitly in appendix \ref{sec:extreme-kerr-initial}.  These
functions depend on a free parameter $J$, the angular momentum of the
data. 
Set
\begin{equation}
  \label{eq:1}
v=v_0+\vv , \quad Y=Y_0+\vY.  
\end{equation} 
To simplify the notation we will write $\vh\equiv (\vv,\vY)$, $u\equiv
(v, Y)$ and $u_0\equiv (v_0, Y_0)$.  Let $\Omega$ be a (unbounded)
domain in $\Rt$. We introduce the following weighted spaces of $C^1$
functions in $\Omega$
\begin{equation}
  \label{eq:3}
\cbn{f} = \sup_{x\in \Omega} \left\{ \sigma^{-\beta} |f|
  +\sigma^{-\beta+1}|\partial f| \right\}, 
\end{equation}
with $\beta < -1/2$ and $\sigma=\sqrt{r^2+1}$, $r=\sqrt{\rho^2+z^2}$. 

Let $\rho_0>0$ be a constant and $K_{\rho_0}$ be the cylinder
$\rho\geq \rho_0$ in $\Rt$. We define the domain $\dy$ by
$\dy=\Rt\setminus K_{\rho_0}$.  The perturbation $y$ is assumed to
vanish in $K_{\rho_0}$. This is consistent with the physical
requirement that the perturbations keep fixed the angular momentum $J$
(see equation \eqref{eq:74} and also \cite{Dain05c} \cite{Dain05e}).
The Banach space $\bs$ is defined by
\begin{equation}
  \label{eq:34}
\bn{\vh}\equiv ||\vv||_{C^1_{\beta}(\Rt)}+
||\vY||_{C^1_{\beta}(\dy)}.
\end{equation}
We consider $\mf$ as a functional on $\mf: \bs \rightarrow \mathbb{R}$.
The following is the main result of  this article.
\begin{theorem}
 \label{t1}
The functional $\mf: \bs \to \mathbb{R}$ defined by
  \eqref{eq:5c} has a strict local minimum at $u_0$.
  That is, there exist $\epsilon>0$ such that
\begin{equation}
  \label{eq:5}
\mf(u_0+\vh)> \mf(u_0),
\end{equation}
for all $\vh\in \bs$ with $\bn{\vh}<\epsilon$ and $ \vh\neq 0$.
\end{theorem}
For simplicity we have assumed that $\vY$ vanished at $K_{\rho_0}$,
the theorem is expected to be true if we impose appropriate decay
conditions for $\vY$ at the axis, however the analysis in this case is
more complicated. We also expect that $u_0$ is an absolute global
minimum of $\mf$ in $\bs$ (see \cite{Dain05c}).

We discuss now the application of theorem \ref{t1} to the proof of 
inequality \eqref{eq:14a} in a neighborhood of extreme Kerr data. What
follows is a summary of the results presented in \cite{Dain05c}. 
A maximal initial data set for Einstein's vacuum equations consists in a
 Riemannian metric $\tilde{h}_{ab}$, and a trace-free symmetric
tensor field $\tilde{K}^{ab}$ such that the vacuum constraint
equations
\begin{align}
 \label{const1}
 \tilde D_b \tilde K^{ab} = 0,\\
 \label{const2}
 \tilde R -\tilde K_{ab} \tilde K^{ab}=0,
\end{align}
are satisfied; where $\tilde{D}_a$ and $\tilde R$ are the Levi-Civita
connection and the Ricci scalar associated with $\tilde{h}_{ab}$. In
these equations the indexes are moved with the metric $\tilde h_{ab}$
and its inverse $\tilde h^{ab}$.

We will assume that the initial data are axially symmetric, that is,
there exist an axial Killing vector $\eta^a$ such that
 \begin{equation}
  \label{eq:8b}
 \pounds_\eta \tilde h_{ab}=0, \quad  \pounds_\eta \tilde K_{ab}=0,
\end{equation}
where $\pounds$ denotes the Lie derivative. 

The Killing vector $\eta^a$ is assumed to be hypersurface orthogonal.
Under these conditions, the metric $\tilde h_{ab}$ can be
characterized by two functions $q,v$ of the coordinates
$\rho,z$. These functions are specified as follow. We
write the metric in the form
\begin{equation}
  \label{eq:1b}
  \tilde h_{ab}=e^{v}h_{ab},
\end{equation}
where the conformal metric $h_{ab}$ is given by 
\begin{equation}
  \label{eq:105}
  h= e^{-2q}(d\rho^2+dz^2)+\rho^2d\phi^2.
\end{equation} 
In these coordinates we have $\eta^a= (\partial/\partial \varphi)^a$
and its  norm with respect to the metric $\tilde h_{ab}$ will be denoted by
$X$ 
\begin{equation}
  \label{eq:72}
X =\eta^a\eta^b\tilde h_{ab}=e^v\rho^2.
\end{equation}

The function $q$ is assumed to be smooth with respect to the
coordinates $\rho,z$. At the axis we impose the regularity
condition
\begin{equation}
  \label{eq:9b}
  q(\rho=0,z)=0,
\end{equation}
and at infinity we assume the following fall-off
\begin{equation}
  \label{eq:10b}
  q=o(r^{-1}), \quad q_{,r}=o(r^{-2}).
\end{equation}

The potential $Y$ is calculated in terms of the second fundamental
form. Define the vector $\tilde S^a$ by 
\begin{equation}
  \label{eq:2t}
  \tilde S_a=\tilde K_{ab}\eta^b-X^{-1}\tilde \eta_a \tilde
  K_{bc}\eta^b\eta^c, 
\end{equation} 
where $\tilde \eta_a=\tilde h_{ab}\eta^b$.  Using equations
\eqref{const1}, \eqref{eq:8b}  and the Killing equation we obtain
\begin{equation}
  \label{eq:73}
\tilde D_{[b}K_{a]}=0, \quad K_a= \tilde \epsilon_{abc}\tilde S^b \eta^c,
\end{equation}
 where $\tilde \epsilon_{abc}$
is the volume element of $\tilde h_{ab}$.
Then there exist a scalar function $Y$ such that $K_a=\tilde D_aY$. 
This function contains the angular momentum $J$ of the data
\begin{equation}
  \label{eq:74}
  J=\frac{1}{8}\left (Y(\rho=0,-z) -Y(\rho=0, z)\right ), \quad z\neq 0.
\end{equation} 

For any data $(\tilde{h}_{ab},\tilde{K}^{ab})$ which satisfy these
assumptions, we have a pair $v,Y$ and its corresponding angular
momentum $J$. Let $u_0\equiv (v_0,Y_0)$ the extreme Kerr data with the
same $J$ and define $\vh\equiv (\vv,\vY)$ by \eqref{eq:1}. We have the
following result.
\begin{corollary} 
  Let $(\tilde{h}_{ab},\tilde{K}^{ab})$ be a maximal, axisymmetric,
  vacuum, initial data with mass $m$ and angular momentum $J$, such
  that the metric satisfies \eqref{eq:1b}--\eqref{eq:105} and
  \eqref{eq:9b}--\eqref{eq:10b}. Define $\vh$ as above. Then, there
  exists $\epsilon>0$ such that for $\bn{\vh}<\epsilon$ the
  inequality \eqref{eq:14a} holds. Moreover, $m=\sqrt{J}$ in this
  neighborhood if and only if the data are the extreme Kerr data.
\end{corollary}
\begin{proof}
  In \cite{Dain05c} it has been proved that under the present
  assumptions we have
\begin{equation}
  \label{eq:75}
m\geq \mf(u_0+\vh).
\end{equation}
By theorem \ref{t1} we have
\begin{equation}
  \label{eq:76}
m\geq \mf(u_0+\vh) >\mf(u_0),
\end{equation}
for  $\vh\neq0$. 
Since $u_0$ is the extreme Kerr data we have $\mf(u_0)=\sqrt{|J|}$
(see\cite{Dain05c}), then the conclusion follows. 
\end{proof}
Finally, we mention  that these results are similar in spirit to the
local proof of the positivity of the mass given in \cite{Brill68}
\cite{Choquet-Bruhat76}. 

\section{Preliminaries}
\label{sec:preliminaries} 
Let us assume $\vh\in \bs$ and consider the real-valued function
\begin{equation}
  \label{eq:19}
i_{\vh}(t)= \mf(u_0+t \vh).
\end{equation}
The  first derivative of $i_{\vh}(t)$ with respect to $t$ is given by
\begin{equation}
  \label{eq:43}
i'_\vh(t)= \frac{1}{16\pi}\int_{\mathbb{R}^3}
  \left\{\partial v \partial \vv + 
\left ( \partial Y \partial \vY -\vv  |\partial Y|^2 \right)X^{-2}\right\} \dv,
\end{equation}
where the $t$ dependence in the right-hand side of \eqref{eq:43} is
encoded in the functions $v,Y,X$ defined by
\begin{equation}
  \label{eq:35}
v\equiv v(t)=v_0+t\vv,\quad Y\equiv Y(t)=Y_0+ t\vY,
\end{equation}
and
\begin{equation}
  \label{eq:33}
X\equiv X(t)=\rho^2e^{v(t)}, \quad X_0 \equiv X(0)=\rho^2e^{v_0}.  
\end{equation}
The second derivative is given by
\begin{equation}
  \label{eq:44}
    i''_\vh(t)= \frac{1}{16\pi}\int_{\mathbb{R}^3}
  \left\{|\partial  \vv |^2 +
\left(2\vv^2|\partial Y|^2
 - 4\vv  \partial Y \partial \vY + |\partial \vY|^2
\right)X^{-2}\right\} \dv.  
\end{equation}
The integrand of the functional $\mf$  is singular at $\rho=0$. However, we
have defined the space $\bs$ only for functions $y$ with support in
$\dy$. Then, the domain of integration of the terms in which $\partial
y$ appears is in fact $\dy$ and hence the integrand is clearly regular
for those terms. Moreover, we have the following. 

\begin{lemma}
\label{l1}
Let $\vh\in\bs$, then the function $i_{\vh}(t)$ is $C^2$ in the $t$
variable.
\end{lemma}
\begin{proof}
  This is a straightforward computation. We will show the the third
  derivative exists for all $t$.  From equation \eqref{eq:44}
  we compute
\begin{equation}
  \label{eq:17}
    i'''_\vh(t)= \frac{1}{8\pi}\int_{\mathbb{R}^3}
\left(-2\vv^3|\partial Y|^2
 + 6\vv^2  \partial Y \partial \vY -3\vv |\partial \vY|^2
\right)X^{-2}  \dv.
\end{equation} 
The terms which contain $\partial \vY$ in \eqref{eq:17} have support in
$\dy$, using inequalities \eqref{eq:49} and \eqref{eq:38} they can be
easily bounded by the norm $\bs$. The only term which does not contain
$\partial \vY$ is given by
\begin{equation}
  \label{eq:30}
  -2\vv^3|\partial Y_0|^2.
\end{equation}
To bound this term we use inequality \eqref{eq:38}.
\end{proof}

In \cite{Dain05c} we have proved that the extreme Kerr initial data is
a critical point of $\mf$, that is we have
\begin{equation}
  \label{eq:39}
i'_{\vh}(0)=0, \quad \text{ for all }\vh\in \bs. 
\end{equation}

In the next section, to prove that the second derivative is coercive
we will need the following auxiliary Hilbert space $\hs$, which is
defined in terms of the weighted Sobolev norms
\begin{align}
  \label{eq:6}
||\vv||^2_{\hs_1} &=  \int_{\Rt}|\partial \vv|^2 \dv    +
  \int_{\Rt} |\vv|^2 r^{-2} \dv , \\
  \label{eq:4}
||\vY||^2_{\hs_2} &=  \int_{\dy}|\partial \vY|^2 \rho^{-4}\dv   + 
  \int_{\dy} |\vY|^2 \rho^{-6} \dv ,
\end{align}
and its corresponding scalar products.  Note that both the domain of
integration and the weight functions are different in these
definitions. The norm in $\hs$ is defined by
\begin{equation}
  \label{eq:41}
\hn{\vh}^2\equiv ||\vv||^2_{\hs_1} + ||\vY||^2_{\hs_2}, 
\end{equation}
with its corresponding scalar product. We have $\bs\subset \hs$.
 
The following weighted Poincare inequalities will be important in the
proof of lemma  \ref{l2}. 
\begin{lemma}
\label{poincare}
Let $\vh\in\hs$ and $\delta\neq 0$ a real number.  Then
\begin{align}
  \label{eq:40}
 |\delta|^{-1} \int_{\Rt}|\partial \vv|^2 r^{-2\delta-1} \dv & \geq  \int_{\Rt}
  |\vv|^2
  r^{-2\delta-3} \dv,\\
 |\delta|^{-1} \int_{\dy}|\partial \vY|^2 \rho^{-2\delta}\dv &\geq  \int_{\dy}
  |\vY|^2 \rho^{-2\delta -2} \dv.\label{eq:40b}
\end{align}
\begin{proof}
  Inequality \eqref{eq:40} is proved in \cite{Bartnik86} (Theorem
  1.3). Using a similar argument we will prove \eqref{eq:40b}. It is
  sufficient to consider functions $y$ with compact support in $\dy$. The key
  is the following identity
\begin{equation}
  \label{eq:55}
\Delta(\ln\rho)=0, \text{ in } \dy, 
\end{equation}
where $\Delta$ is the Laplacian in $\Rt$. 
Testing this equation with $\rho^{-2\delta}\vY^2$ gives
\begin{equation}
  \label{eq:59}
\int_{\dy} \partial(\rho^{-2\delta}y^2)\partial(\ln\rho) \dv=0.
\end{equation}
Which expands to 
\begin{equation}
  \label{eq:60}
\int_{\dy} \rho^{-2\delta-2}y^2 \leq |\delta|^{-1}\int_{\dy}
\rho^{-2\delta-1}|y| |\partial_\rho y| \dv,
\end{equation}
and H\"olders inequality gives \eqref{eq:40b}. 
\end{proof}

\end{lemma}
The following lemma shows that
$i''_{\vh}(t)$ is uniformly continuous with respect to the $\hs$ norm.
\begin{lemma}
\label{l2}
Let $0<t<1$. For every $\epsilon>0$ there exist $\eta(\epsilon)$ such that for 
$ \bn{\vh} <\eta(\epsilon)$
we have
\begin{equation}
  \label{eq:7}
|i''_{\vh}(t) - i''_{\vh}(0)| \leq \epsilon \hn{\vh}^2. 
\end{equation}
\end{lemma}
\begin{proof}
The proof is a straightforward but tedious calculation. 
  Using \eqref{eq:44} we calculate
\begin{equation}
  \label{eq:36}
  i''_{\vh}(t) - i''_{\vh}(0)=\int_{\Rt} (A + B+ D )\dv,
\end{equation}
where
\begin{align}
  \label{eq:37}
A & = 2\vv^2 \left(\frac{|\partial Y|^2}{X^2}-\frac{|\partial
    Y_0|^2}{X_0^2}   \right),\\
B &= -4\vv\partial\vY \left( \frac{\partial Y}{X^2}-\frac{\partial
    Y_0}{X_0^2}   \right),\\
D &= |\partial \vY |^2 \left( X^{-2}-X_0^{-2}   \right).
\end{align}
We decompose further $A$ and $B$ in the following way
\begin{equation}
  \label{eq:42}
A=A_1+A_2+A_3, \quad B=B_1+B_2,
\end{equation}
where
\begin{align}
  \label{eq:45}
A_1 &= 2\vv^2 \frac{|\partial Y_0|^2}{X^2_0}(e^{-2t\vv}-1),\\
A_2 &= 4t \frac{\vv^2}{X^2_0}e^{-2t\vv}\partial Y_0 \partial \vY,\\
A_3 &=  2t^2 \frac{\vv^2}{X^2_0}e^{-2t\vv} |\partial \vY|^2,
\end{align}
and
\begin{align}
  \label{eq:46}
B_1 & = -4 (e^{-2t\vv}-1)\vv X^{-2}_0\partial \vY \partial Y,\\
B_2 & = -4t \vv X^{-2}_0|\partial \vY|^2.
\end{align}
In the following, we will bound each individual term. We will
repeatedly use the inequalities \eqref{eq:38}--\eqref{eq:64} given in
appendix \ref{sec:extreme-kerr-initial} for the extreme Kerr initial data. 

Note that $A_1$ is the only term without $\partial \vY$, and hence the
only term with support in $\Rt$, the other terms have support in
$\dy$. To bound $A_1$ we use inequality \eqref{eq:38}  
\begin{equation}
  \label{eq:47}
\int_{\Rt} A_1 \dv \leq 2C_1 (e^{2\eta}-1)\int_{\Rt} \vv^2X_0^{-2} \dv
\leq 2C_1 (e^{2\eta}-1) \hn{\vh},
\end{equation} 
where we have used
\begin{equation}
  \label{eq:48}
|\vv|\leq\sigma^\beta ||\vv||_{C^1_{\beta}(\Rt)} \leq
||\vv||_{C^1_{\beta}(\Rt)} \leq \bn{\vh}\leq \eta.  
\end{equation}
To bound the other terms we will use inequality \eqref{eq:49}. For
$A_2$ we have 
\begin{align}
  \label{eq:50}
\int_{\dy} A_2 \dv & \leq 4e^{2\eta} \eta \int_{\dy} \vv \frac{\partial
  Y}{X_0}\frac{\partial y}{X_0}\dv,\\
&\leq   4e^{2\eta} \eta \left( \int_{\dy} \vv^2
\frac{|\partial  Y|^2}{X^2_0}\dv \right)^{1/2} \left( \int_{\dy}
\frac{|\partial  y|^2}{X^2_0}\dv\right)^{1/2},\\
& \leq 4 C_1 e^{2\eta} \eta \hn{\vh}^2,       
\end{align}
where in the second line we have used H\"older inequality and in the
third line inequalities \eqref{eq:38} and  \eqref{eq:49}.

The terms $A_3$ is simpler
\begin{equation}
  \label{eq:51}
\int_{\dy} A_3 \dv\leq 2 e^{2\eta}\eta^2  \hn{\vh}^2.
\end{equation}
In the similar way we get
\begin{align}
  \label{eq:52}
\int_{\dy} B_1 \dv & \leq 4 (e^{2\eta}-1) \left( \int_{\dy} \vv^2
\frac{|\partial  Y|^2}{X^2_0}\dv \right)^{1/2} \left( \int_{\dy}
\frac{|\partial  y|^2}{X^2_0}\dv\right)^{1/2}\\
&\leq (e^{2\eta}-1) C_1 \hn{\vh}^2.
\end{align}
For $B_2$ we get
\begin{equation}
  \label{eq:53}
\int_{\dy} B_2 \dv  \leq 4 \eta  \hn{\vh}^2. 
\end{equation}
Finally for $D$ we have
\begin{equation}
  \label{eq:54}
\int_{\dy} D \dv\leq (e^{2\eta}-1)   \hn{\vh}^2.
\end{equation}
Since all the coefficients multiplying $\hn{\vh}^2$ are smooth with
respect to $\eta$ and go to zero as $\eta$ goes to zero we get
\eqref{eq:7}.
\end{proof}

\section{Positivity of the second variation}
\label{sec:second-variation}
From equation \eqref{eq:44}, it is far from obvious that the second
variation evaluated at the critical point $u_0$ is positive definite.
In order to prove that, 
the key ingredient is the following remarkable identity proved by
Carter \cite{Carter71}. In terms of our variables it has the following form
\begin{equation}
  \label{eq:9}
F + \vv G'_{v}+\vY G'_{Y}+2\vv\vY G_{Y} - X^{-2}\vY^2 G_{v} =
\partial \left(\vv\partial \vv +\vY X^{-1}    \partial \left(  \vY
    X^{-1} \right)\right),   
\end{equation}
 where
\begin{align}
  \label{eq:2}
G_v(t) & = \Delta v + X^{-2}  |\partial Y|^2 ,\\
G_Y(t) &= \partial( X^{-2}\partial Y),
\end{align}
the derivatives with respect to $t$ of these functions are given
\begin{align}
  \label{eq:10}
G'_v(t) & =\Delta \vv + \left( 2\partial \vY \partial Y -2\vv  \partial_a
Y\partial^a Y  \right)X^{-2},\\
G'_Y(t) &= \partial (X^{-2} \left(\partial \vY-2\vv \partial Y\right) ),
\end{align}
and the positive definite function $F$ is given by 
\begin{equation}
  \label{eq:26}
F(t)=\left(\partial\vv+ \vY X^{-2} \partial Y \right )^2 +
\left(\partial( \vY  X^{-1})- X^{-1} \vv\partial Y\right)^2
+\left( X^{-1} \vv\partial Y  - \vY X^{-2} \partial X\right)^2
\end{equation}
The identity \eqref{eq:9} is valid for arbitrary $v,Y,\vv,\vY$ and it
is straightforward to check.   

For $\vh \in \bs$, integrating by part we obtain 
\begin{equation}
  \label{eq:11}
-\int_{\Rt} \left( \vv G'_{v}+\vY G'_{Y}\right) \dv =
16\pi i''_{\vh}(t). 
\end{equation}  

Equation \eqref{eq:39} is equivalent to  $G_{v}(0)=G_{Y}(0)=0$. If we
integrate the identity \eqref{eq:9} in $\Rt$ for $t=0$, the boundary terms in
the right-hand side 
vanishes for all $\vh \in \bs$, then we get
\begin{equation}
  \label{eq:27}
i''_{\vh}(0)=\int_{\Rt} F(0) \dv \geq 0.
\end{equation}
Moreover, $i''_{\vh}(0)$  is strictly positive. Assume that we have
$i''_{\vh}(0)= 0$ for some $\vh  \in \bs$. Then, $F(0)=0$.  We 
rewrite the second and the third term in \eqref{eq:26} in  the following form
\begin{equation}
  \label{eq:13}
\left(X\partial \vYb + \vYb \partial X  - X^{-1}
  \vv\partial Y\right)^2 
+\left( X^{-1} \vv\partial Y  - \vYb  \partial
  X\right)^2\geq 2^{-1} X^2  (\partial \vYb )^2,
\end{equation}
where
\begin{equation}
  \label{eq:32}
\bar y = \frac{y}{X^2}.
\end{equation}
From this we deduce that $F(0)= 0$ implies $ \partial (\vY
X_0^{-2})=0$, and hence, by the assumption $\vh\in \bs$, we obtain
$\vY=0$. Then we deduce $\vv=0$ from the first term in \eqref{eq:9}.

Equation \eqref{eq:27} is a necessary condition for a local minimum, however it is not
sufficient. The following lemma gives the sufficient condition used in
the proof of theorem \ref{t1}. Note that in order to formulate this
coercive condition we need the auxiliary Hilbert space $\hs$ defined in
the previous section. 
\begin{lemma}
\label{l3}
  There exists $\lambda>0$ such that for all $\vh\in \bs$ we have
\begin{equation}
  \label{eq:8}
i''_{\vh}(0)\geq \lambda \hn{\vh}^2.
\end{equation}
\end{lemma}
\begin{proof}
Let $\vh\in \bs$. Note that  $i''_\vh(0)$, as function of $\vh$, defines a
bilinear form $a(\vh,\vh)$ given by
\begin{equation}
  \label{eq:24}
a(\vh,\vh) \equiv i''_\vh(0)= \int_{\Rt} F(0)\dv,
\end{equation}
where $F$ is given explicitly by \eqref{eq:26}.

Inequality \eqref{eq:8} is equivalent to the following variational
problem 
\begin{equation}
  \label{eq:15}
\lambda = \inf_{\vh\in \bs} \frac{a(\vh,\vh)}{\hn{\vh}^2}.
\end{equation}
Since  $a(\gamma \vh,\gamma \vh)=\gamma^2a(\vh,\vh) $ for every real number
$\gamma$, this variational problem  is equivalent to 
\begin{equation}
  \label{eq:15b}
\lambda = \inf_{\vh\in M} a(\vh,\vh),
\end{equation}
where
\begin{equation}
  \label{eq:18}
 M= \{\vh\in \bs: \hn{\vh}^2=1 \}. 
\end{equation}
It is clear that
$\lambda\geq0$ because $a(\vh,\vh)$ is positive definite.  We
want to prove that $\lambda >0$. 

Let us assume that $\lambda=0$. Then,  there exists a sequence
  $\{\vh_n\}$ such that 
\begin{equation}
  \label{eq:16}
\hn{\vh_n}=1,\quad \text{ for all } n,
\end{equation}
and
\begin{equation}
  \label{eq:14}
\lim_{n\to \infty}a(\vh_n,\vh_n)= 0. 
\end{equation}
By the inequality \eqref{eq:13} we have,
\begin{equation}
  \label{eq:25}
\lim_{n\to \infty} a(\vh_n,\vh_n) \geq2^{-1} \int_{\dy}X_0^2 (\partial \vYb_n)^2=0.
\end{equation}
Using the bounds \eqref{eq:49}, \eqref{eq:49b} and the Poincare
inequality \eqref{eq:40b} we have
\begin{align}
  \label{eq:62}
\int_{\dy}X_0^2 (\partial \vYb_n)^2\dv & \geq \int_{\dy}\rho^4
(\partial \vYb_n)^2\dv\\ 
&\geq 2 \int_{\dy}\rho^2
 |\vYb_n|^2\dv\\
& \geq  \frac{2}{C^2_2}\int_{\dy}\rho^{-6} |\vY_n|^2\dv.
\end{align}
Then,
\begin{equation}
  \label{eq:63}
\lim_{n\to \infty} \int_{\dy}\rho^{-6} |\vY_n|^2\dv =0.
\end{equation} 
Replacing $\vYb_n$ by $X_0^{-2}\vY_n$  in \eqref{eq:25} we get the inequality
\begin{equation}
  \label{eq:65}
a(\vh_n,\vh_n) + \frac{1}{4} \int_{\dy}X_0^{-2} |\vY_n|^2
|\partial \ln X_0 |^2\geq 4\int_{\dy}X_0^{-2} (\partial \vY_n)^2. 
\end{equation} 
From this inequality, using the bounds \eqref{eq:49}, \eqref{eq:49b}
and \eqref{eq:64} we obtain
\begin{equation}
  \label{eq:66}
a(\vh_n,\vh_n) + \frac{C_3}{4} \int_{\dy} \rho^{-6} |\vY_n|^2
\geq 4C_2 \int_{\dy}\rho^{-4} (\partial \vY_n)^2.
\end{equation}
We apply \eqref{eq:63} and  \eqref{eq:14}  to get
\begin{equation}
  \label{eq:67}
\lim_{n\to \infty} \int_{\dy}\rho^{-4} |\partial \vY_n|^2\dv =0,
\end{equation}
and then we have
\begin{equation}
  \label{eq:28}
\lim_{n\to \infty} ||\vY_n||_{\hs_2}=0.
\end{equation}

In the previous inequalities, we have only used the second and the
third terms of $F$, the first term implies
\begin{equation}
  \label{eq:29}
\int_{\dy} \vY_n^2X_0^{-4}
|\partial Y_0|^2\dv + a(\vh_n,\vh_n) \geq 2^{-1}\int_{\Rt} |\partial\vv_n|^2.
\end{equation}
We have
\begin{equation}
  \label{eq:68}
\int_{\dy} \vY_n^2X_0^{-4}
|\partial Y_0|^2\dv \leq  C_1 \int_{\dy} \vY_n^2r^{-2} \rho^{-4}\leq
C_1\int_{\dy} \vY_n^2 \rho^{-6}\dv,
\end{equation}
where in the last inequality we have used that $r\geq \rho$. 

From  \eqref{eq:68}, \eqref{eq:29} and \eqref{eq:63} we deduce 
\begin{equation}
  \label{eq:69}
\lim_{n\to \infty} \int_{\Rt}  |\partial \vv_n|^2\dv =0,
\end{equation}
and using the Poincare inequality \eqref{eq:40} we finally obtain
\begin{equation}
  \label{eq:70}
\lim_{n\to \infty} ||\vv_n||_{\hs_1}=0.
\end{equation}

Equations \eqref{eq:70} and \eqref{eq:28} contradict \eqref{eq:16},
hence we have that $\lambda >0$.      
\end{proof}

\section{Proof of theorem \ref{t1}}
\label{sec:proof-theorem-reft1}

\begin{proof}
  
  The proof follows standard arguments, see for example
  \cite{Giaquinta96} chapter 5 and \cite{Zeidler85} chapter 40.
  
  In lemma \ref{l1} we have proved that the function $i_\vh(t)$ is $C^2$
  with respect to $t$. The classical Taylor theorem  for $t=1$ yields
\begin{equation}
  \label{eq:20}
\mf(u_0+ \vh)-\mf(u_0)=i_\vh(1)-i_\vh(0)=\frac{i''_\vh(\vartheta)}{2}, \quad
0<\vartheta<1.
\end{equation}
To prove \eqref{eq:5} we will show that $i''_\vh(\vartheta)\geq 0$ and
$i''_\vh(\vartheta)= 0\Rightarrow \vh=0$.

From lemma \ref{l2} we have that for every $\epsilon>0$ there exist
$\eta(\epsilon)$ such that the following inequality holds
\begin{equation}
  \label{eq:21}
|i''_\vh(\vartheta)-i''_\vh(0)|\leq \epsilon\hn{\vh}^2, 
\end{equation}
for every $\bn{\vh}<\eta(\epsilon)$. From this inequality we deduce
\begin{equation}
  \label{eq:22}
i''_\vh(0)- \epsilon\hn{\vh}^2\leq i''_h(\vartheta).
\end{equation}
We use lemma \ref{l3} and inequality \eqref{eq:22} to conclude that
\begin{equation}
  \label{eq:23}
(\lambda- \epsilon)\hn{\vh}^2\leq i''_\vh(\vartheta).
\end{equation}
Choosing $\eta(\epsilon)$ such that $\epsilon<\lambda$ the desired
result follows.
\end{proof}

\section*{Acknowledgments}
It is a pleasure to thank R. Bartnik, R. Beig, L. Simon and R. Schoen
for valuable discussions. 
Special thanks to  Todd
Oliynyk for illuminating discussions concerning  Carter identity. 

This work has been supported by the Sonderforschungsbereich SFB/TR7 of
the Deutsche Forschungsgemeinschaft.
 
\appendix
\section{Extreme Kerr initial data}\label{sec:extreme-kerr-initial}

Consider the extreme Kerr metric, with angular momentum $J$, in
Boyer-Lindquist coordinates $(t,\tilde r,\theta,\phi)$. The
corresponding potentials $X_0,Y_0$ are given by
\begin{equation}
  \label{eq:56}
  Y_0 = 2J(\cos^3\theta-3\cos\theta)-
  \frac{2J^2\cos\theta\sin^4\theta}{\Sigma},  
\end{equation}
and 
\begin{align}
  \label{eq:57}
X_0 & =\left(\frac{(\tilde r^2+|J|)^2 -r^2 |J|
      \sin^2\theta}{\Sigma} \right)\sin^2\theta,\\
\label{eq:57b}
&= \left(\tilde r^2+|J| +\frac{2|J|^{3/2} \tilde r \sin^2\theta
  }{\Sigma} \right)\sin^2\theta, 
\end{align}
where
\begin{equation}
  \label{eq:58}
\tilde r = r+\sqrt{|J|}, \quad 
\Sigma=\tilde r^2+|J| \cos^2 \theta,
\end{equation}
The function $v_0$ is given by the definition \eqref{eq:33}
\begin{equation}
  \label{eq:12}
v_0=\ln X_0 -2\ln{\rho},
\end{equation}
where $\rho=r\sin\theta$.

From \eqref{eq:56} and \eqref{eq:57} we get the  following inequality 
\begin{equation}
  \label{eq:38}
\frac{|\partial Y_0|^2}{X_0^2}\leq C_1 r^{-2} \quad \text{in } \Rt.  
\end{equation}
From equation \eqref{eq:57b} we obtain a lower bound for $X_0$
\begin{equation}
  \label{eq:49}
\rho^2 \leq X_0  \quad \text{in } \Rt.
\end{equation}
To get an upper bound we use equation  \eqref{eq:57} to obtain
\begin{equation}
  \label{eq:61}
X^2_0\leq 16\rho^4 \left(1+\frac{\sqrt{|J|}}{r} \right)^4.
\end{equation}
From this equation we deduce
\begin{equation}
  \label{eq:49b}
X^2_0\leq C_2\rho^4  \quad \text{in } \dy,
\end{equation}
where
\begin{equation}
  \label{eq:71}
 C_2 =16\rho^4 \left(1+\frac{\sqrt{|J|}}{\rho_0} \right)^4.
\end{equation}
Note that  $\lim_{\rho_0\to 0} C_2=\infty$.

From equation \eqref{eq:57b} and \eqref{eq:12} we obtain
\begin{equation}
  \label{eq:31}
|\partial v_0|^2\leq C'_3r^{-2}\quad  \text{in } \Rt,
\end{equation}
where $C'_3>0$ and then
\begin{equation}
  \label{eq:64}
|\partial \ln X_0|^2\leq  |\partial v_0|^2+4\rho^{-2}\leq
C_3\rho^{-2}\quad \text{in } \Rt, 
\end{equation}
where $C_3=C'_3+4$ and  we have used that $r\geq \rho$.

\end{document}